\documentclass[11pt,]{article}
\usepackage{lmodern}
\usepackage{amssymb,amsmath}
\usepackage{ifxetex,ifluatex}
\usepackage{fixltx2e} % provides \textsubscript
\ifnum 0\ifxetex 1\fi\ifluatex 1\fi=0 % if pdftex
  \usepackage[T1]{fontenc}
  \usepackage[utf8]{inputenc}
\else % if luatex or xelatex
  \ifxetex
    \usepackage{mathspec}
  \else
    \usepackage{fontspec}
  \fi
  \defaultfontfeatures{Ligatures=TeX,Scale=MatchLowercase}
\fi
\IfFileExists{upquote.sty}{\usepackage{upquote}}{}
\IfFileExists{microtype.sty}{%
\usepackage{microtype}
\UseMicrotypeSet[protrusion]{basicmath} % disable protrusion for tt fonts
}{}
\usepackage[margin=1in]{geometry}
\usepackage{hyperref}
\hypersetup{unicode=true,
            pdftitle={Imputing Missing Values with External Data},
            pdfauthor={Karolinska Institutet},
            pdfborder={0 0 0},
            breaklinks=true}
\usepackage{natbib}
\usepackage{graphicx,grffile}
\makeatletter
\def\maxwidth{\ifdim\Gin@nat@width>\linewidth\linewidth\else\Gin@nat@width\fi}
\def\maxheight{\ifdim\Gin@nat@height>\textheight\textheight\else\Gin@nat@height\fi}
\makeatother
\setkeys{Gin}{width=\maxwidth,height=\maxheight,keepaspectratio}
\IfFileExists{parskip.sty}{%
\usepackage{parskip}
}{% else
\setlength{\parindent}{0pt}
\setlength{\parskip}{6pt plus 2pt minus 1pt}
}
\providecommand{\tightlist}{%
  \setlength{\itemsep}{0pt}\setlength{\parskip}{0pt}}
\ifx\paragraph\undefined\else
\let\oldparagraph\paragraph
\renewcommand{\paragraph}[1]{\oldparagraph{#1}\mbox{}}
\fi
\ifx\subparagraph\undefined\else
\let\oldsubparagraph\subparagraph
\renewcommand{\subparagraph}[1]{\oldsubparagraph{#1}\mbox{}}
\fi

\let\rmarkdownfootnote\footnote%
\def\footnote{\protect\rmarkdownfootnote}

\usepackage{titling}

\newcommand{\subtitle}[1]{
  \posttitle{
    \begin{center}\large#1\end{center}
    }
}

  \title{Imputing Missing Values with External Data}
  \pretitle{\vspace{\droptitle}\centering\huge}
  \posttitle{\par}
\subtitle{Robert Thiesmeier, Matteo Bottai, Nicola Orsini}
  \author{Karolinska Institutet}
  \preauthor{\centering\large\emph}
  \postauthor{\par}
  \predate{\centering\large\emph}
  \postdate{\par}
  \date{\href{mailto:robert.thiesmeier@ki.se}{\nolinkurl{robert.thiesmeier@ki.se}}}

\begin{document}
\maketitle
\begin{abstract}
Missing data is a common challenge across scientific disciplines. Current imputation methods require the availability of individual data to impute missing values. Often, however, missingness requires using external data for the imputation. In this paper, we introduce a new Stata command, \textbf{mi impute from}, designed to impute missing values using linear predictors and their related covariance matrix from imputation models estimated in one or multiple external studies. This allows for the imputation of any missing values without sharing individual data between studies. We describe the underlying method and present the syntax of \textbf{mi impute from} alongside practical examples of missing data in collaborative research projects.
\end{abstract}

\begin{center}
Draft v1: \today
\end{center}

\section{Introduction}
Missing data is a common challenge to statistical inference in a variety of scientific disciplines \citep{CarpenterKenward2013}. Commonly, sporadic missing data occurs in single studies and can be handled with conventional imputation approaches available with the \texttt{mi} suite of commands in Stata. First, \texttt{mi impute} is used to create imputed datasets by fitting an imputation model for missing variables. Second, \texttt{mi estimate} is employed to fit a substantive model across imputed datasets and combine estimates with Rubin's rules \citep{StataMI2023}. Current imputation approaches require individual-level data to be used as the basis for imputation, but often borrowing information from other studies is necessary to impute missing values in a single study. For example, many collaborative research efforts involving multiple studies contain systematically missing data (100\% missing values) for key variables in one or more studies. While there has been extensive effort to address systematically missing data in individual participant data meta-analysis with multiple imputation (MI) \citep{Burgess2013, Resche-Rigon2013, Jolani2015, Quartagno2016, Audigier2018, Resche-Rigon2018}, a crucial prerequisite for current MI approaches is the availability of individual data. Often though sharing individual data is not possible because of logistic or legal barriers. Consequently, standard imputation procedures become infeasible as there is no basis for imputation. While a few studies have proposed and described the idea of MI across studies without the need to share individual data \citep{Secrest2020, Shu2023}, software facilitating imputation without individual data is currently not available. In this paper, we introduce a new command - \texttt{mi impute from} - designed to handle univariate imputation of missing values using external data. The key feature of this approach is to fit an imputation model in an external dataset that has some information on the variable that is missing. Instead of sharing individual data, the regression coefficients of an imputation model are shared between studies and used as the basis of imputation.\\
\noindent The remaining paper is structured as follows. First, the methods underlying \texttt{mi impute from} are presented with different imputation options including quantile, multinomial, and binary logistic imputation. Second, the structure and options of the command are shown. Third, \texttt{mi impute from} is illustrated with different simulated examples including imputing missing values for a confounding variable, an effect modifier, and a predictor in collaborative research projects. We conclude by highlighting the potential of the new command and pointing out future developments.

%%%%%%%%%%%%%%%%%%%%%%%%%%%%%%%%%%%%
\section{Methods} \label{sec:methods}
In this section, we outline the general steps of imputing missing values using external data. Here, we consider systematically missing data, as this is a common missing data challenge involving multiple studies. Such missing values may affect important covariates that are not measured or collected. Although the focus is on systematically missing data, \texttt{mi impute from} can be used for any type of missing data and currently allows to impute continuous, discrete, and binary variables with quantile, multinomial, and binary logistic regression, respectively.
%%%%%%%%%%%%%%%%%%%%%%%%%%%%%%%%%%%%
\subsection{Imputation of missing values using external data}
Let $z_{ij}$ denote observations from independent random variables $Z_{ij}$ in the $j$-th study. The index $i$, identifying an individual, ranges from 1 to $n_j$ and the index $j$, identifying a study, ranges from 1 to $J$. In addition, let us denote with $\mathbf{c}_{ij}=(c_{1i}, c_{2i}, ..., c_{ki})$ a $k$-dimensional set of predictors of $z_{ij}$. Suppose the vector $\mathbf{c}_{ij}$ is completely observed for all $i \in {1, . . . , n_j}$ and all $j \in {1, . . . , J}$, while $z_{ij}$ is missing in the $j$-th study. Let $M_j \subset \{1, \ldots, n_j\}$ be the set of indexes corresponding to the individuals with missing values for $z_{ij}$ in the $j$-th study. The cardinality of $M_j$ divided by the sample size $n_j$ denotes the proportion of missing data for the $j$-th study, that is $FM_j = |M_j|/n_j$, ranging from 0 (no missing) to 1 (all missing, i.e., systematically missing). Systematically missing data in a single study is present if $FM_j=1$. For the remainder of this paper, let $J = A \cup B$, with $A = {j \in J : FM_j = 1}$, denote the set of studies with systematically missing data, whereas the set of studies $B = {j \in J : FM_j < 1 }$ are used as the basis of imputation.\\
First, in each study $J \in B$, fit an imputation model on $z_{i}$ dependent on a set of predictors $\mathbf{c}_{i}$. An imputation model is fitted in the form of either a quantile, multinomial logistic, or logistic regression model. The resulting vector of imputation regression coefficients $\boldsymbol{\hat{\gamma}}_i = (\gamma_{0i}, \gamma_{1i}, \gamma_{2i}, ...,\gamma_{ki})$ and the related covariance matrix $\mathbf{\hat{\Phi}}_i$ are saved to be shared with other studies in $J \in A$. \\
Second, the files including the regression coefficients $\boldsymbol{\hat{\gamma}}_i$ and covariances $\mathbf{\hat{\Phi}}_i$ are exported into files (e.g., Excel or delimited text files) that can be shared with other studies. In $J \in A$, the files are imported and transformed to matrices to be used for the imputation of missing data with \texttt{mi impute from}. The command \texttt{mi\_impute\_from\_get} facilitates this step. If \texttt{mi\_impute\_from\_get} recognises multiple input files, a series of inverse-variance weighted least squares models are estimated to compute $\boldsymbol{\bar{\gamma}}_i(\mathbf{c}_{i})$ and $\mathbf{\bar{\Phi}}_i$. Regardless of the number of files, a random number from a normal distribution is drawn and further used as the regression coefficient $\boldsymbol{\gamma}_i^*(\mathbf{c}_{i}) \sim \mathcal{N}(\hat{\boldsymbol{\gamma}}_i(\mathbf{c}_{i}), \widehat{SE}(\hat{\boldsymbol{\gamma}}_i(\mathbf{c}_{i})))$ to be used for the $m$-th imputation of the missing value $z_{i}^{(m)}$.\\
Third, the command \texttt{mi\_impute\_from\_get} returns formatted matrices that are used for the imputation of impute missing values on $z_{i}$ in $J \in A$. Next, \texttt{mi impute from} executes $m = 1, 2, ..., M$ imputations of $z_{i}$ conditional on the predictors $\mathbf{c}_{i}$. The key feature of \texttt{mi impute from} is that missing values are imputed \textit{without} individual data as the basis for the imputation. One can impute continuous, discrete, and binary missing variables with quantile, multinomial, and binary logistic regression, respectively. The individual assignment of imputed values to the missing data for each of the three imputation methods is explained in Section \ref{imp_methods}.\\
Last, a substantive model can be fit in each imputed data set and estimates are pooled with Rubin's rules, provided by \texttt{mi estimate}.
%%%%%%%%%%%%%%%%%%%%%%%%%%%%%%%%%%%%%%%%%%%%%%%%%%%%%%%%%%%%%%%%%%%%%%%%%%%
\subsection{Imputation methods}\label{imp_methods}
Three imputation methods are currently implemented for continuous, discrete, and binary variables that are handled with \texttt{qreg}, \texttt{mlogit}, and \texttt{logit}, respectively. 
%%%%%%%%%%%%%%%%%%%%%%%%%%%%%
\subsubsection{Continuous variables}
A continuous missing variable can be imputed by sharing the estimates of linear predictors of quantile regression models which has been previously described in \citet{Bottai2013} and \citet{Thiesmeier2024}.\\
In $J \in B$, the cumulative distribution function of $z_{i}$ conditionally on a set of predictors $\mathbf{c}_{i}$ can be derived by estimating the $p$-quantile of $z_{i}$ with a quantile regression model where $p$ ranges from 0.01 to 0.99 in steps of 0.01
\begin{equation}
\hat{Q}_{z_{i}|\mathbf{c}_{i}}(p) = \mathbf{c}_{i}\boldsymbol{\gamma}_i(p)
\quad  p \in \{0.01, 0.02, \ldots, 0.99\}
\label{eq:imp_model}
\end{equation}
\noindent There are 99 sets of regression coefficients $\hat{\boldsymbol{\gamma}}_i(p)$ defining the cumulative distribution function for any linear combination of predictors. Each missing value $z_{i}^{(m)}$ is replaced by $M$ independent imputed values. In $J \in A$, a random draw $U_{i}$ from a continuous uniform distribution $\mathcal{U}(0,1)$ is taken. The $m$-th imputation $z^{(m)}_{i}$ for the $i$-th individual is computed as the weighted average of the $F$ and $F+1$ conditional predicted quantiles
\begin{equation}
    z^{(m)}_{i} = (1-\bmod) \cdot \hat Q_{z_{i}|\mathbf{c}_{i}}(F) + \bmod \cdot \hat Q_{z_{i}|\mathbf{c}_{i}}(F+1)
\end{equation}
where $F = \lfloor U_{i}\%\rfloor$ and $\bmod=U_{i}\%-\lfloor U\%\rfloor$.
%%%%%%%%%%%%%%%%%%%%%%%%%%%%%%%%%%%%
\subsubsection{Categorical variables}
A categorical missing variable $z_i$ with $K$ levels can be imputed by sharing $k$ linear predictors of multinomial logistic regression models. We denote $\theta_{ik}$ as the probability that $z_{i}$ is equal to level $k$, given a set of predictors $\mathbf{c}_{i}$, $\theta_{ik} | \mathbf{c}_{i} = P(z_{i} = k | \mathbf{c}_{i})$ with $\sum_{k=1}^K \theta_{ik} = 1$. In $J \in B$, a multinomial logistic regression imputation model is specified as 
\begin{equation}
    \ln\left(\frac{\theta_{ik}}{\theta_{ir}} \right) = \boldsymbol{\gamma}_{ik}(\mathbf{c}_{i})
    \label{mlogit}
\end{equation}
where $r$ is the reference level. The estimated predicted conditional probabilities of falling into the levels of the categorical variable $z_{i}$ are denoted as $\hat{\theta}_{ik} = \hat{P}(z=i|\mathbf{c}_{i})$ for any individual in $J \in B$. In the set of studies $J \in A$, $\hat{\theta}_k = \hat{P}(z=i|\mathbf{c}_{i})$ is used to compute the conditional predicted probabilities of the missing variable 
\begin{equation}
\hat{\theta}_{ik} = 
\begin{cases}
\frac{1}{1 + \sum_{k=2}^K e^{\hat{\boldsymbol{\gamma}}_{ik}(\mathbf{c}_{i})}} & \text{if } k=r \\[12pt] 
\frac{e^{\hat{\boldsymbol{\gamma}}_{ik}(\mathbf{c}_{i})}}{1 + \sum_{k=2}^K e^{\hat{\boldsymbol{\gamma}}_{ik}(\mathbf{c}_{i})}} & \text{if } k>r
\end{cases}
\end{equation}
The conditional predicted cumulative distribution function, $\hat{\Theta}_{ik}$, is equal to the sum of the probabilities of the missing variable being less than or equal to $k$:
\begin{equation}
\hat{\Theta}_{ik} = \hat{P}(z_{i} \leq k | \mathbf{c}_{i}) = \sum_{k=1}^K \hat{\theta}_{ik}
\end{equation}
In $J \in A$, a random draw $U_{i}$ from a continuous uniform distribution $\mathcal{U}(0,1)$ is taken. The $m$-th imputation $z^{(m)}_{i}$ for the $i$-th individual is obtained by mapping $U_{i}$ to the predicted cumulative probabilities $\hat{\Theta}_{ik}$.
%%%%%%%%%%%%%%%%%%%%%%%%%%%%%
\subsubsection{Binary variables}
A binary (coded as 0/1) missing variable can be imputed by sharing the linear predictors of logistic regression models. When $z_{i}$ is binary, we denote $\theta_i$ as the probability that $z_{i} = 1$, given a set of predictors $\mathbf{c}_{i}$, $\theta_i | \mathbf{c}_{i} = P(z_{i} = 1 | \mathbf{c}_{i})$. In $J \in B$, we specify a logistic regression imputation model in the form of
\begin{equation}
    \ln\left(\frac{\theta_i}{1 - \theta_i} \right) = \boldsymbol{\gamma}_i(\mathbf{c}_{i})
\end{equation}
Here, $\hat{\boldsymbol{\gamma}}_i(\mathbf{c}_{i})$ is the estimated linear predictor. The predicted conditional probability can be expressed as 
\begin{equation}
    \hat{\theta}_i = \frac{e^{\hat{\boldsymbol{\gamma}}_i(\mathbf{c}_{i})}}{1 + e^{\hat{\boldsymbol{\gamma}}_i(\mathbf{c}_{i})}}
\end{equation}
In the set of studies $J \in A$, we take a random draw $U_i$ from a continuous uniform distribution $U(0,1)$ and assign $z_{i}^{(m)} = 1$ if $U_{i} > \hat{\theta}_i$ and 0 otherwise.
%%%%%%%%%%%%%%%%%%%%%%%%%%%%%
\section{The syntax}
\subsection{mi impute from}
\texttt{mi impute from} fills in missing values using an imputation model estimated in one or multiple external studies. The command assumes that variables used in the imputation model are available in the current data.
\begin{verbatim}
    mi impute from ivar [if] [, b(matname) v(matname) imodel(string)]
\end{verbatim}
After \texttt{mi impute from}, \texttt{mi estimate} can be used to compute estimates of regression coefficients by fitting an estimation command to $M$ imputed data sets.

\subsubsection{Options}
\begin{itemize}
\tightlist
    \item \texttt{b(matname)} specifies a vector of regression coefficients of an imputation model for \textit{ivar}.
    \item \texttt{v({matname})} specifies a matrix of covariances for the imputation model to be used for \textit{ivar}.
    \item \texttt{imodel} specifies the imputation model, where \textit{string} can be either \texttt{qreg}, \texttt{mlogit}, or \texttt{logit}.
    \item \texttt{qreg} specifies that the matrix \texttt{b(\textit{matname})} contains $99 (q=0.01(.01)0.99)$ linear predictors of the quantitative variable \textit{ivar} to be imputed.
    \item \texttt{mlogit} the imputation model is a multinomial logistic regression model for the categorical variable \textit{ivar}.
    \item \texttt{logit} the imputation model is a logistic regression model for the binary variable \textit{ivar}.
\end{itemize}
Additionally, the options \texttt{add()}, \texttt{replace}, \texttt{rseed()}, \texttt{double} are available, same as in \texttt{[MI] mi impute}.

\subsubsection{Saved results}
\texttt{mi impute from} stores the following in \texttt{r()}:

\textbf{Scalars}\\
\begin{itemize}
\tightlist
    \item \texttt{r(M)} number of observations
    \item \texttt{r(M\_add)} number of added imputations
    \item \texttt{r(M\_update)} number of updated imputations
    \item \texttt{r(k\_ivars)} number of imputed variables (always \textbf{1})
    \item \texttt{}{r(pp)} \textbf{1} if perfect prediction detected, \textbf{0} otherwise
    \item \texttt{r(N\_g)} number of imputed groups (\textbf{1} if \textbf{by()} is not specified)
\end{itemize}\vspace{0.2cm}

\textbf{Macros}\\
\tightlist
\begin{itemize}
    \item \texttt{r(method)} name of imputation method (\textbf{from})
    \item \texttt{r(ivars)} names of imputation variables
    \item \texttt{r(ngstate)} random-number state used
\end{itemize}\vspace{0.2cm}

\textbf{Matrices}\\
\begin{itemize}
\tightlist
    \item \texttt{r(N)} number of observations in imputation sample in each group
    \item \texttt{r(N\_complete)} number of complete observations in imputation sample in each group
    \item \texttt{r(N\_incomplete)} number of incomplete observations in imputation sample in each group
    \item \texttt{r(N\_imputed)} number of imputed observations in imputation sample in each group
\end{itemize}
%%%%%%%%%%%%%%%%%%%%%%%%%%%%%
\subsection{mi\_impute\_from\_get}
The command \texttt{mi\_impute\_from\_get} facilitates the use of external imputation models by reading the files and formatting matrices to be passed to  \texttt{mi\_impute\_from}. If multiple files are specified, \texttt{{mi\_impute\_from\_get}} combines regression coefficients across files using an inverse-variance weighted least squares model.

\begin{verbatim}

    mi_impute_from_get [, options]
    
\end{verbatim}

\subsubsection{Options}
\begin{itemize}
\tightlist
    \item \texttt{b(\text{filename} [\text{filename} [...]])} specifies a list of files containing the estimated regression coefficients for the imputation model for \textit{ivar}.
    \item \texttt{v(\text{filename} [\text{filename} [...]])} specifies a list of files containing the estimated covariances of the regression coefficients for the imputation model for \textit{ivar}.
    \item \texttt{colnames(string)} specifies a list of variable names (including the constant \texttt{\_cons}) included in the linear predictor of the imputation model.
    \item \texttt{tf(string)} format of the files (either tab delimited .txt or excel .xlsx) containing the estimated imputation model. Adding the format (.txt, .xlsx) to each file name in \texttt{b()} and \texttt{v()} is not needed. The default format for all the files is tab delimited (.txt).
    \item \texttt{imodel(string)} the type of imputation model (\texttt{qreg}, \texttt{mlogit}, \texttt{logit}) passed in \texttt{b()}. Of note, for a quantitative variable one has to pass the linear predictor for $99 (q=0.01(.01).99)$ quantiles of the imputation model. A loop can be used to compute all the estimated linear predictors in a single file. For a categorical variable with $k$ levels, $k$ linear predictors of multinomial logistic regression models must be passed. The reference level is recognized by the fact that all regression coefficients are equal to zero.
    \item \texttt{values(\textit{numlist}} specifies all the numerical values of the categorical variable to be imputed. This is needed when using \texttt{imodel(mlogit)} as the imputation model.
    \item \texttt{path(string)} specifies the location of the file names specified in options \texttt{b()} and \texttt{v()}.
\end{itemize}

\subsubsection{Saved results}
\texttt{mi\_impute\_from\_get} stores the following in \texttt{r()}:

\textbf{Matrices}\\
\begin{itemize}
    \item \texttt{r(get\_ib)} regression coefficients for the imputation model
    \item \texttt{r(get\_iV)} covariance matrix for the imputation model
\end{itemize}
%%%%%%%%%%%%%%%%%%%%%%%%%%%%%%%%%%%%%%%%%%
\section{Examples}
The following section demonstrates multiple examples to impute missing values with external data. The individual data used for the examples are available as supplementary material. The command can be downloaded from the SSC Archive typing: \texttt{ssc install mi\_impute\_from}. See Section \ref{dataav} to access the data sets and Stata code to replicate the examples.
%%%%%%%%%%%%%%%%%%%%%%%%%%%%%%%%%%%%%%%%%%
\subsection{Example 1: Missing confounder}
Considering multiple studies to answer the same research question can improve the generalisability of findings and is increasingly becoming the norm in international research collaborations \citep{Toh2013}. However, data between studies often cannot be shared for legal barriers \citep{casaletto2023}. In this example, we consider a collaborative research project involving five observational studies. The aim is to estimate the adjusted odds ratio for the exposure in all five studies. All studies have collected data on the outcome, $Y$, the exposure, $X$, and and two confounding variables, $C$ and $Z$. However, in Study 1 the confounding variable $Z$ is 100\% missing and it is not possible to estimate the fully-adjusted odds ratio for the exposure. Either Study 1 has to be excluded from the fully-adjusted analysis or the adjustment for the confounder $Z$ has to be disregarded. Both options are undesirable. We therefore borrow information from studies that have collected data on the confounder $Z$ to impute the missing values in Study 1. Table \ref{table_1} shows descriptive characteristics and estimated odds ratios of the five studies.

\begin{table}[h!]
\centering
\small
  \begin{tabular}{lccccc}
    \hline
    & \textbf{Study 1} & \textbf{Study 2} & \textbf{Study 3} & \textbf{Study 4} & \textbf{Study 5}\\
    & N=6,437 & N=1,334 & N=8,623 & N=3,603 & N=6,673\\
    \hline
    Outcome (\%) & 12.0 & 24.1 & 30.1 & 15.3 & 39.8 \\
    Exposure (\%) & 20.5 & 9.8 & 18.8 & 18.5 & 18.5 \\
    Confounder $C$ (\%) & 39.9 & 37.5 & 28.3 & 25.6 & 23.4 \\
    {Confounder $Z$ (Median)}  & \textbf{NA} & 0.4 & 0.5 & 0.4 & 0.5 \\
    \hspace{0.3cm}(Q25,Q95) & \textbf{NA} & (0.1, 4.1) & (0.1, 3.9) & (0.1, 3.7) & (0.1, 3.9) \\ 
    \\
    \hline
    Crude OR & 2.1 (1.8, 2.5)& 2.1 (1.4, 3.1) & 2.4 (2.1, 2.7) & 1.8 (1.4, 2.2) & 2.0 (1.8, 2.3) \\
    $C$ Adjusted OR & 1.6 (1.4, 1.9)& 1.7 (1.2, 2.6) & 1.8 (1.6, 2.1) & 1.5 (1.2, 1.9) & 1.7 (1.5, 1.9) \\
    $Z$ \& $C$ Adjusted OR & \textbf{NA} & 1.6 (1.1, 2.4) & 1.6 (1.5, 1.8) & 1.4 (1.2, 1.8) & 1.4 (1.2, 1.6) \\
    \hline
  \end{tabular}
  \caption{Descriptive statistics and estimated odds ratios (95\% Confidence Interval (CI)) with a different degree of adjustment. Q25 = $25^{th}$ percentile, Q95 = $95^{th}$ percentile.}
  \label{table_1}
\end{table}

\subsubsection{Using a single external study}
First, one can choose a single study to specify an imputation model for the confounder $Z$. For example, let us estimate 99 conditional quantiles for $Z$ as a function of $Y, X, \text{and }C$ in Study 2. The imputation regression coefficients are exported into a text file that is easy to share between studies. In Study 1, the text files of the regression coefficients (\texttt{b\_study2}) and their covariances (\texttt{v\_study2}) from the imputation model are imported with the command \texttt{mi\_impute\_from\_get}. The resulting matrices for the regression coefficients (\texttt{get\_ib}) and the covariance matrix (\texttt{get\_iV}) are now used for the imputation of the missing values for the confounder $Z$ with \texttt{mi\_impute\_from}. The correct order and number of variables used in the imputation model is specified in the option \texttt{colnames}. The missing values for $Z$ are imputed 10 times.

\begin{verbatim}

. quietly use study_1, clear 
. mi set wide
. mi register imputed z
. mi_impute_from_get , b(b_study2) v(v_study2) colnames(y x c _cons) imodel(qreg) 
. mat ib = r(get_ib)
. mat iV = r(get_iV)
. mi impute from z , add(10) b(ib) v(iV) imodel(qreg) rseed(24092024)

External imputation using qreg              Imputations =       10
User method from                                  added =       10
Imputed: m=1 through m=10                       updated =        0

------------------------------------------------------------------
                   |               Observations per m             
                   |----------------------------------------------
          Variable |   Complete   Incomplete   Imputed |     Total
-------------------+-----------------------------------+----------
                 z |          0         6437      6437 |      6437
------------------------------------------------------------------
(Complete + Incomplete = Total; Imputed is the minimum across m
 of the number of filled-in observations.)
 
\end{verbatim}
The total number of imputed values equals the total sample size in Study 1. The command \texttt{mi impute from} has imputed 100\% missing values for the confounder $Z$ using external regression coefficients. Finally, \texttt{mi estimate} is used to fit a multivariate logistic regression model across imputed data sets and combine the results with Rubin's rules to obtain an estimate for the fully-adjusted odds ratio in Study 1.

\begin{verbatim}

. mi estimate, post eform noheader : logit y x c z

------------------------------------------------------------------------------
           y |     exp(b)   Std. err.      t    P>|t|     [95% conf. interval]
-------------+----------------------------------------------------------------
           x |   1.477021   .1348225     4.27   0.000     1.234993    1.766482
           c |   3.090796   .2632513    13.25   0.000     2.615522    3.652433
           z |   1.317996   .0529826     6.87   0.000     1.213572    1.431406
       _cons |   .0516561   .0042865   -35.71   0.000     .0438569    .0608423
------------------------------------------------------------------------------

\end{verbatim}

The estimated fully-adjusted odds ratio in Study 1 is OR = 1.5 (95\% CI = 1.2, 1.8) and we are able to include Study 1 in the collaborative research project considering \textit{all} five studies.

\subsubsection{Using multiple external studies}
Alternatively to using a single study, one can derive an imputation model from all four studies. After specifying an imputation on $Z$ conditional on $Y, X, \text{and } C$ in Study 2, 3, 4, and 5, we import the list of text files of the regression coefficients (\texttt{b\_study2 b\_study3 b\_study4 b\_study5}) and covariance matrices (\texttt{v\_study2 v\_study3 v\_study4 v\_study5}) from all four studies with \texttt{mi\_impute\_from\_get}. The command recognises multiple files and computes a weighted average of the imputation regression coefficients resulting in the average regression coefficients and variances in matrices \texttt{ib} and \texttt{iV}.

\begin{verbatim}

. quietly use study_1, clear 
. mi set wide
. mi register imputed z
. mi_impute_from_get , ///
>         b(b_study2 b_study3 b_study4 b_study5) ///
>         v(v_study2 v_study3 v_study4 v_study5) ///
>         colnames(y x c _cons) imodel(qreg) 
. mat ib = r(get_ib)
. mat iV = r(get_iV)
. mi impute from z , add(10) b(ib) v(iV) imodel(qreg) rseed(24092024)

External imputation using qreg              Imputations =       10
User method from                                  added =       10
Imputed: m=1 through m=10                       updated =        0

------------------------------------------------------------------
                   |               Observations per m             
                   |----------------------------------------------
          Variable |   Complete   Incomplete   Imputed |     Total
-------------------+-----------------------------------+----------
                 z |          0         6437      6437 |      6437
------------------------------------------------------------------
(Complete + Incomplete = Total; Imputed is the minimum across m
 of the number of filled-in observations.)

\end{verbatim}

A multivariate logistic regression model is fitted in the imputed data sets and results are combined with Rubin's rules to obtain a final estimate of the fully-adjusted odds ratio in Study 1. 

\begin{verbatim}

. mi estimate, post eform noheader : logit y x c z

------------------------------------------------------------------------------
           y |     exp(b)   Std. err.      t    P>|t|     [95% conf. interval]
-------------+----------------------------------------------------------------
           x |   1.319035   .1257386     2.90   0.004      1.09403    1.590317
           c |   3.258008   .2790564    13.79   0.000     2.754424    3.853662
           z |   1.375821   .0629058     6.98   0.000     1.252929    1.510767
       _cons |   .0506925   .0043144   -35.04   0.000     .0428502    .0599701
------------------------------------------------------------------------------

\end{verbatim}

The estimated imputed fully-adjusted odds ratio is 1.3 with a 95\% CI from 1.1 to 1.6. The next step in a collaborative effort would be a meta-analytical approach to combine the fully-adjusted odds ratios across studies.\\
Of note, all studies were simulated under a heterogeneous data generating mechanism. The fully-adjusted odds ratio in Study 1 before the confounder $Z$ was set to missing is OR = 1.3 (95\% CI = 1.1, 1.5). The underlying value of the fully-adjusted confounding effect in all studies was set to OR = 1.4. Using multiple studies as the basis for imputation resulted in an odds ratio close to the estimated value before systematically missing data were introduced. This example illustrates how \texttt{mi impute from} can lead to the inclusion of the fully-adjusted odds ratio in Study 1 to the overall collaborative effort.

\subsection{Example 2: Missing effect modifier}
Multiple trials are often considered to study differential treatment effects \citep{riley2010}. In practice, the effect of a treatment commonly depends on different levels of a third factor, such as disease stage \citep{Riley2020}. The rationale for a collaborative effort involving multiple trials is then to increase the possibility to detect such differential treatment effects and increase the generalisability of the findings. In this example, we consider two large randomised controlled trials with a randomly allocated treatment, $X$, and an important discrete effect-modifier, $Z$, with three levels (low, medium, high). The outcome was defined as time-to-death from all causes. It is hypothesised that the treatment has a beneficial, null, and harmful effect at the low, medium, and high level of the effect modifier, respectively. Therefore, the target of statistical inference is the effect of the treatment at the three levels of $Z$. In Trial 1, however, the effect modifier $Z$ has not been measured. Trial 2 has observed data on $Z$ which can be used to specify an imputation model. Table \ref{table_2} shows the characteristics of the two trials.

\begin{table}[h!]
\centering
  \begin{tabular}{lcc}
    \hline
    & \textbf{Trial 1} & \textbf{Trial 2} \\
    & N=4,000 & N=3,175 \\
     \hline
     \\
    Number of cases & 1,893 & 2,074 \\
    Rate (per 1000)& 64.9 & 107.5 \\
    $\text{HR}_{X|Z=0}$ (95\% CI) & \textbf{NA}  & 0.5 (0.4, 0.6) \\
    $\text{HR}_{X|Z=1}$ (95\% CI) & \textbf{NA}  & 0.8 (0.7, 0.9) \\
    $\text{HR}_{X|Z=2}$ (95\% CI) & \textbf{NA} & 1.5 (1.3, 1.7) \\
    \hline
  \end{tabular}
  \caption{Characteristics of Trial 1 and 2, including the number of cases, the baseline failure rate per 1000 person-years, and the estimated Hazard ratio (HR) (95\% Confidence Interval (CI)) of the treatment $X$ at the low, medium, and high level of the effect modifier $Z$.}
  \label{table_2}
\end{table}

To be able to include both trials in the collaborative effort, we specify a multinomial logistic regression model in Trial 2 and export the imputation regression coefficients to Trial 1. The imputation model must be specified congenial to the outcome model \citep{meng1994, white2011}. As such, interaction terms between 1) the treatment and the cumulative hazard  (\texttt{x\_cumh}), and 2) the treatment and indicator variable for death (\texttt{x\_d}) are included in the imputation model. In \texttt{mi\_impute\_from\_get}, the three levels of the effect modifier $Z$ are specified in \texttt{values(0 1 2)}. In addition, we model the interaction effects of the treatment $X$ and the effect modifier $Z$ with indicator variables. As such, the survival model estimating the effect of the treatment $X$ at the low, medium, and high level of the effect modifier $Z$ is obtained with the main effect (\texttt{x}) and two product terms (\texttt{x\_zi1}, and \texttt{x\_zi2}). Before missing values for $Z$ in Trial 2 are imputed, all indicator variables are registered as passive imputation variables. Finally, the missing values of the effect modifier $Z$ are imputed 10 times.

\begin{verbatim}

. quietly use study_1, clear 
. mi set wide
. qui mi stset time, fail(death)
. mi register regular cumh _d x_cumh x_d
(variable _d already registered as regular)
. mi register imputed z
. mi register passive zi1 zi2 x_zi1 x_zi2
. mi_impute_from_get , ///
>         b(b_study2) v(v_study2) ///
>         colnames(x cumh _d x_cumh x_d _cons) values(0 1 2) imodel(mlogit) 
. mat ib = r(get_ib)
. mat iV = r(get_iV)
. mi impute from z , add(10) b(ib) v(iV) imodel(mlogit) rseed(24092024)

External imputation using mlogit            Imputations =       10
User method from                                  added =       10
Imputed: m=1 through m=10                       updated =        0

------------------------------------------------------------------
                   |               Observations per m             
                   |----------------------------------------------
          Variable |   Complete   Incomplete   Imputed |     Total
-------------------+-----------------------------------+----------
                 z |          0         4000      4000 |      4000
------------------------------------------------------------------
(Complete + Incomplete = Total; Imputed is the minimum across m
 of the number of filled-in observations.)

\end{verbatim}

After imputing the missing values for the effect modifier $Z$, a Cox regression model is fitted in each imputed data set and the estimates are pooled with Rubin's rules.

\begin{verbatim}
. mi estimate , post eform noheader saving(miestfile, replace):  ///
>         stcox x zi1 zi2 x_zi1 x_zi2 

------------------------------------------------------------------------------
          _t |     exp(b)   Std. err.      t    P>|t|     [95% conf. interval]
-------------+----------------------------------------------------------------
           x |   .5358579   .1577109    -2.12   0.056     .2815539    1.019853
         zi1 |   3.521413   .6111299     7.25   0.000     2.429803    5.103438
         zi2 |   2.082912   .3135047     4.88   0.000      1.51759    2.858825
       x_zi1 |   1.455193   .5192257     1.05   0.314     .6669019    3.175259
       x_zi2 |   3.029879   .9916384     3.39   0.004     1.501813    6.112724
------------------------------------------------------------------------------

\end{verbatim}

In Trial 1, the effect of the treatment is estimated to reduce the mortality rate by 50\% (HR = 0.5; 95\% CI = 0.3, 1.0) and 20\% (HR = 0.8, 95\% CI = 0.6, 1.1) in the low and medium level of the effect modifier $Z$, respectively. On the contrary, the effect of the treatment in the high level of the effect modifier $Z$ conferred a 60\% higher mortality rate (HR = 1.6; 95\% CI = 1.1, 2.3).\\
Of note, the set values of the trials were $\text{HR}_{X|Z=0} = 0.5$,  $\text{HR}_{X|Z=1}$ = 1, and $\text{HR}_{X|Z=2}$ = 1.5. Additionally, before $Z$ was set to 100\% missing in Trial 1, the estimated HR were $\text{HR}_{X|Z=0}$ = 0.5 (95\% CI = 0.4, 0.6),  $\text{HR}_{X|Z=1}$ = 1 (95\% CI = 0.9, 1.1), and $\text{HR}_{X|Z=2}$ = 1.7 (95\% CI = 1.4, 2.0). We therefore know that we have not introduced substantial bias in the imputed estimates of the HR in Trial 1 and were able to include both trials in the overall analysis.

\subsection{Example 3: Missing predictor}
Developing prediction models based on large observational studies is common in many health-related research fields \citep{debray2017}. In this example, we consider a large study in which a prediction model for the outcome $Y$ is specified with all available predictors, say $X$ and $C$. The predictive capacity of the model is quantified as the Area Under the Curve (AUC) and is estimated at AUC = 0.662. In another study, Study 2, a similar prediction model has been estimated. However, additionally, Study 2 has collected data on a strong binary predictor, $Z$, and the discriminative ability of the model is estimated at AUC = 0.821. The investigator of Study 1 would like to evaluate to what extent the predictor $Z$ from Study 2 may or may not improve the AUC. To understand the potential impact of including the predictor $Z$, let us use information on $Z$ from Study 2. Table \ref{table_3} shows the characteristics of both studies including the AUC. 

\begin{table}[h!]
\centering
  \begin{tabular}{lcc}
    \hline
    & \textbf{Study 1} & \textbf{Study 2} \\
    & N=24,087 & N=23,614 \\
     \hline
     \\
    AUC \textit{without} predictor $Z$ & 0.662 & 0.715 \\
    AUC \textit{with} predictor $Z$ & \textbf{NA} & 0.821 \\
    \hline
  \end{tabular}
  \caption{Sample size and the Area Under the Curve (AUC) of the prediction model in Study 1 and 2.}
  \label{table_3}
\end{table}

In Study 2, a logistic regression model is specified on $Z$ conditional on $Y, X, \text{and } C$. The regression coefficients and covariance matrix are exported as text files to Study 1. The text files are then used for \texttt{mi\_impute\_from\_get}. The missing predictor $Z$ is imputed 10 times. 

\begin{verbatim}

. quietly use study_1, clear 
. mi set wide
. mi register imputed z
. mi_impute_from_get , ///
>                 b(b_study2) v(v_study2) ///
>                 colnames(y x c _cons) imodel(logit) 
. mat ib = r(get_ib)
. mat iV = r(get_iV)
. mi impute from z , add(10) b(ib) v(iV) imodel(logit) rseed(24092024)

External imputation using logit             Imputations =       10
User method from                                  added =       10
Imputed: m=1 through m=10                       updated =        0

------------------------------------------------------------------
                   |               Observations per m             
                   |----------------------------------------------
          Variable |   Complete   Incomplete   Imputed |     Total
-------------------+-----------------------------------+----------
                 z |          0        24087     24087 |     24087
------------------------------------------------------------------
(Complete + Incomplete = Total; Imputed is the minimum across m
 of the number of filled-in observations.)

\end{verbatim}

\begin{figure}[h!] 
    \centering
    \includegraphics[width=\textwidth]{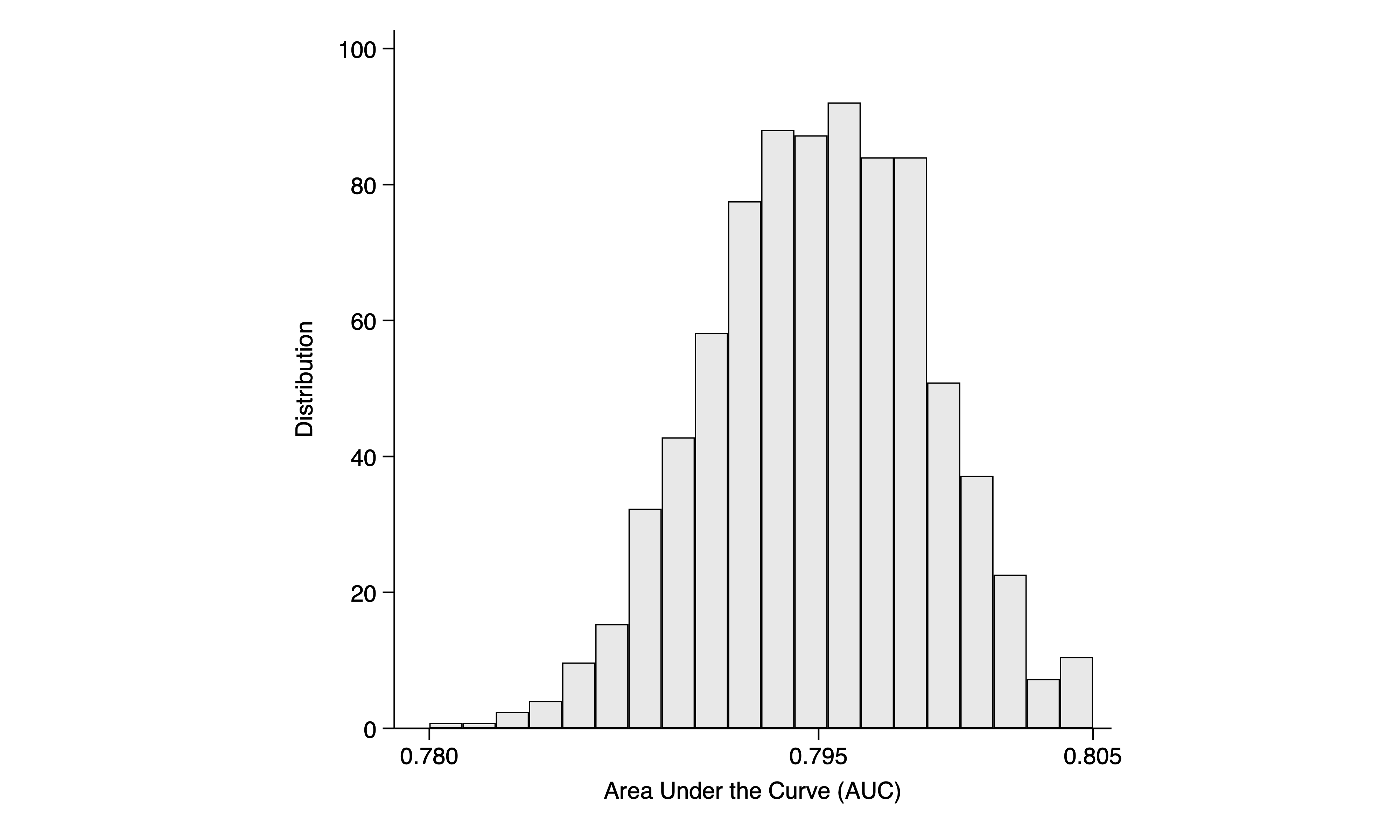}
    \caption{Distribution of the Area Under the Curve of 1000 imputation of the predictor $Z$ in Study 1.}
    \label{figure1} 
\end{figure}

A multivariate logistic regression model is fitted in each imputed data set to obtain 10 estimates for the AUC. A final estimate of the AUC can be obtained as an average of the estimates across imputed data sets. With 10 imputations, the AUC of the prediction model in Study 1 increased from 0.662 to 0.796 after including the predictor $Z$. In Figure \ref{figure1}, the distribution of the AUC is shown after adding 1000 imputations. Across imputations, the AUC is centered around the average value of 0.795, ranging from 0.780 to 0.805.

This example illustrated how \texttt{mi impute from} can be used for the design of a prediction model by borrowing information from another study to evaluate the improvement of the discriminative ability in the presence of an additional predictor.
%%%%%%%%%%%%%%%%%%%%%%%%%%%%%
\section{Conclusion}
In this paper we have described and illustrated a new command - \texttt{mi impute from} - to impute missing values using external data using imputation regression coefficients instead of individual data as the basis for imputation. The syntax and use of the command were illustrated with three examples imputing missing values for a continuous confounder, discrete effect modifier, and binary predictor.\\
The command \texttt{mi impute from} allows users to use external data to impute missing values. The files including the regression coefficients are easy to share and can circumvent logistic and legal barriers of data protection. Users are able to impute both sporadically and systematically missing data which is a unique feature of \texttt{mi impute from} and is currently not possible with any other imputation commands in Stata. Further, \texttt{mi impute from} allows for flexible imputation of continuous, discrete, and binary variables and can be easily extended to accommodate more imputation methods in the future. Current limitations are the 1) restriction of use for univariate imputation of missing data, and 2) no integration into imputation with chained equation (\texttt{mi impute chained}). A previously conducted simulation study shows robust results of the approach described in \citet{Thiesmeier2024}. However, more investigations are planned into the behaviour of the command in more complex scenarios.\\
In conclusion, \texttt{mi impute from} can be a valuable command to be used for external data imputation, particularly in the setting when sharing of individual data across multiple studies is not feasible.

%%%%%%%%%%%%%%%%%%%%%%%%%%%%%
\section*{Data availability}\label{dataav}
The datasets and Stata code to replicate the examples is available on \url{https://github.com/robertthiesmeier/mi_impute_from}.

\section*{Acknowledgments}
This work was supported by the National Infrastructure NEAR, supported by the Swedish Research Council [grant numbers Dnrs 2017-00639 and 2021-00178].

\section*{About the authors}
Robert Thiesmeier is a doctoral student in biostatistics at the Department of Global Public Health and the Department of Neurobiology, Care Sciences and Society, Karolinska Institutet, Sweden. E-mail: robert.thiesmeier@ki.se
\\ \\
Matteo Bottai is a Professor of Biostatistics at the Unit of Biostatistics, Institute of Environmental Medicine, Karolinska Institutet, Stockholm, Sweden.
\\ \\
Nicola Orsini is an Associate Professor of Medical Statistics at the Department of Global Public Health, Karolinska Institutet, Sweden.

\renewcommand\refname{References}
\bibliography{mi_impute_from_arXiv.bbl}

\end{document}